\documentclass[10pt,a4paper]{article}
\usepackage[T1]{fontenc}
\usepackage{microtype}
\usepackage{amsmath,amssymb}
\usepackage{amsfonts}
\usepackage{mathrsfs}
\usepackage[table]{xcolor}  
\usepackage{booktabs}
\usepackage{array}
\usepackage{multirow}
\usepackage{longtable}
\usepackage{tabularx}
\usepackage{xltabular}

\usepackage{graphicx}
\usepackage[export]{adjustbox}
\usepackage{pdflscape}
\usepackage{rotating}
\usepackage{caption}
\usepackage{subcaption}
\usepackage{authblk}
\usepackage{etoolbox}
\usepackage{orcidlink}
\usepackage{textcomp}
\usepackage[margin=2.5cm,top=1.5cm]{geometry}
\usepackage{hyperref}
\usepackage[numbers,sort&compress]{natbib}
\hypersetup{colorlinks=true, linkcolor=blue, citecolor=blue, urlcolor=blue}
\providecommand{\keywords}[1]{%
  \par\vspace{0.5ex}%
  \noindent\textbf{Keywords: }#1\par
}
\numberwithin{equation}{section}
\numberwithin{figure}{section}
\numberwithin{table}{section}
\newcolumntype{C}{>{\centering\arraybackslash}X}
\newcolumntype{L}{>{\raggedright\arraybackslash}X}
\newcolumntype{R}{>{\raggedleft\arraybackslash}X}
\renewcommand{\arraystretch}{1.3}
\begin{document}

\begin{center}
\LARGE{Repetitive Penrose Process in Rotating 4D Einstein–Gauss–Bonnet Black Holes \\\;\\\;}
\par\end{center}

\begin{center}
{\bf Mohammad Reza Alipour\orcidlink{0000-0001-8074-7865}}\footnote{\bf mohamad.alipour.1994@gmail.com; mr.alipour@stu.umz.ac.ir}\\
{\it School of Physics, Damghan University, P.~O.~Box 3671641167, Damghan, Iran}\\
{\it Center for Theoretical Physics, Khazar University, 41 Mehseti Street, Baku, AZ1096, Azerbaijan}
\end{center}

\begin{center}
{\bf Mohammad Ali S. Afshar\orcidlink{0009-0001-3133-5992}}\footnote{\bf m.a.s.afshar@gmail.com}\\
{\it School of Physics, Damghan University, P.~O.~Box 3671641167, Damghan, Iran}\\
{\it Center for Theoretical Physics, Khazar University, 41 Mehseti Street, Baku, AZ1096, Azerbaijan}
\end{center}

\begin{center}
{\bf Saeed Noori Gashti\orcidlink{0000-0001-7844-2640}}\footnote{\bf sn.gashti@du.ac.ir; saeed.noorigashti70@gmail.com}\\
{\it School of Physics, Damghan University, P.~O.~Box 3671641167, Damghan, Iran}\\
{\it Center for Theoretical Physics, Khazar University, 41 Mehseti Street, Baku, AZ1096, Azerbaijan}
\end{center}

\begin{abstract}
We investigate the repetitive Penrose process for neutral particles in a rotating
four-dimensional Einstein--Gauss--Bonnet black hole obtained through the modified
Newman--Janis algorithm, developing a nonlinear iterative scheme in which the mass,
angular momentum, and irreducible mass are updated after each extraction event. Imposing
the triple turning-point condition, we obtain a closed-form solution of the conservation
equations for energy, angular momentum, and radial momentum that reduces to the Kerr
result in the limit of vanishing coupling. The distinctive feature of this background is
that, although the Gauss--Bonnet coupling $\alpha$ is a fixed constant of the action and
is not carried by the infalling fragments, the dimensionless coupling
$\hat\alpha=\alpha/M^{2}$ grows at every iteration as the mass decreases, so that the
effective Gauss--Bonnet correction is self-amplified along the sequence. We find that a
larger coupling lowers the extremal spin, contracts the ergosphere, and reduces the
number of admissible decays, forbidding the process near the horizon at strong coupling
while permitting it at larger decay radii; the termination is controlled throughout by
the incident particle. The energy return on investment decreases monotonically with
$\hat\alpha$ and the growth of the irreducible mass is suppressed relative to Kerr,
whereas the energy utilization efficiency is non-monotonic: below a critical coupling the
parameter space exhibits a four-region structure with 
a bounded window in which the EGB black hole is more efficient than Kerr; 
this four-region structure collapses into a three-region one above a critical coupling.
This coupling-driven reorganization of the efficiency
landscape has no analogue in the Kerr, Reissner--Nordstr\"om, Kerr--de~Sitter,
accelerating Kerr, or Kerr--Newman cases.
\end{abstract}


\tableofcontents

\section{Introduction}
\label{sec:intro}

The possibility of extracting energy from a rotating black hole was established by Penrose,
who showed that a particle entering the ergosphere and decaying there can send a
negative-energy fragment across the horizon while a second fragment escapes to infinity
with more energy than the original~\cite{Penrose1969,PenroseFloyd1971}. The kinematics of
the mechanism, the bound on the escaping energy, and the constraints on the decay were
subsequently clarified by Bardeen, Press, and Teukolsky and by Wald~\cite{Bardeen1972,Wald1974},
and Denardo and Ruffini extended the idea to charged particles in the
Reissner--Nordstr\"om geometry, where an analogous generalized ergosphere supports
negative-energy states~\cite{Denardo1973,Ruffini1975}.

The energetics of these processes are governed by the irreducible mass. Christodoulou and
Ruffini showed that a black hole possesses a mass-energy formula in which $M_{\mathrm{irr}}$
sets the part of the mass that cannot be extracted by classical means, and that
transformations are reversible when $M_{\mathrm{irr}}$ is held fixed and irreversible when
it grows~\cite{Christodoulou1970,ChristodoulouRuffini1971}. The connection between the
irreducible mass and the horizon area, $A=16\pi M_{\mathrm{irr}}^{2}$, together with
Hawking's area theorem~\cite{Hawking1971} and the identification of the area with
entropy~\cite{Bekenstein1973}, fixes the amount of energy that can be liberated and the
thermodynamic cost of doing so.

For a long time the Penrose process was treated as a single event, and attempts to repeat
it were formulated linearly, with the black-hole parameters held fixed between decays.
Ruffini and collaborators recently pointed out that such a treatment violates energy
conservation, and that a consistent repetitive process must update the mass and angular
momentum after every decay~\cite{Ruffini2025PRL,Ruffini2025PRR}. Once this is done the
sequence becomes strongly nonlinear: most of the lost rotational energy is converted not
into escaping radiation but into irreducible mass, the spin approaches a positive lower
bound, and the process terminates after a finite number of steps with only a small
fraction of the mass extracted.

This framework has since been extended to a variety of backgrounds. Hu, Cai, and
Wang studied the repetitive electric Penrose process in Reissner--Nordstr\"om
spacetime and found that the black-hole charge cannot be driven exactly to
zero through a finite sequence of decays, a behavior they interpreted as a
thermodynamic third-law analogue~\cite{HuCaiWang2026}. In Kerr--de~Sitter
black holes, Wang and Zeng showed that a larger cosmological parameter
enhances both the single-extraction capability and the energy return on
investment~\cite{WangZeng2025}, while Zeng and Wang found that, for
accelerating Kerr black holes, the energy utilization efficiency can exceed
the $50\%$ bound obtained in the Kerr case when the decay occurs sufficiently
close to the horizon~\cite{ZengWang2026}.
The process has also been generalized to charged particles in an initially extremal
Kerr--Newman black hole, where two independent electromagnetic couplings separately
control the incident particle's access to the ergoregion and the depth of the
negative-energy states available to the captured fragment~\cite{Aliporcharged2026}. An
attractive black-hole--fragment coupling enhances both the extraction efficiency and the
energy return on investment and, unlike in the Reissner--Nordstr\"om case, allows the
horizon charge to pass through zero and reverse sign without violating the area theorem
or cosmic censorship. Also, the process has been extended to the Konoplya--Zhidenko rotating non-Kerr metric,
where a purely geometric deformation parameter---unrelated to any physical hair---was
likewise shown to raise both the energy return on investment and the energy utilization
efficiency relative to the Kerr case, particularly at larger decay radii~\cite{Zeng2026KZ}.
In each of these extensions, the
additional spacetime parameter---charge, cosmological constant,
acceleration, or metric deformation---tends to make the repetitive process
either more efficient or easier to sustain than in the Kerr case.

Einstein--Gauss--Bonnet gravity provides a natural setting in which to test how a purely
gravitational modification of the background affects this picture. The theory supplements
General Relativity with the quadratic Gauss--Bonnet invariant, which is topological in four
dimensions but, through a consistent regularization, contributes to the field equations
and leaves an imprint on black-hole solutions controlled by a coupling $\alpha$, with
General Relativity recovered as $\alpha\to0$. A rotating solution can be generated from the
static one by the modified Newman--Janis algorithm~\cite{Kumar2020EGB}, giving a
stationary, axisymmetric, asymptotically Kerr-like geometry whose horizon and ergosphere
depend on $\alpha$. Because the Penrose process is sensitive precisely to the size of the
ergoregion and to the location of the horizon, this spacetime is a suitable arena for
studying how quadratic curvature corrections modify repetitive energy extraction.

In this work we develop the repetitive Penrose process for neutral particles in the
rotating four-dimensional Einstein--Gauss--Bonnet black hole. Imposing the triple
turning-point condition, we obtain a closed-form solution of the conservation laws and
build a nonlinear iterative scheme that updates the mass, angular momentum, and irreducible
mass after every decay. A feature specific to this background is that, although $\alpha$ is
a fixed constant of the action and is not carried by the infalling fragments, the
dimensionless coupling $\hat\alpha=\alpha/M^{2}$ grows as the mass decreases, so that the
effective strength of the correction changes along the sequence and the geometry must be
recomputed at each step. We show that this running reduces the extractable reservoir and
the number of admissible decays, that the termination is set throughout by the incident
particle, and that the utilization efficiency reorganizes into a four-region structure
below a critical coupling and a three-region structure above it, with a bounded window in
which the Einstein--Gauss--Bonnet black hole outperforms Kerr.

The paper is organized as follows. Section~\ref{sec:egb} reviews the rotating
four-dimensional Einstein--Gauss--Bonnet black hole, its horizon structure, and its
ergosphere. Section~\ref{sec:rpp} formulates the repetitive Penrose process in this
spacetime, presents the analytic solution and the iterative scheme, and derives the
critical spin thresholds that terminate the sequence. Section~\ref{sec:numerical-results} contains
the numerical analysis and the classification of the efficiency regimes, and
Sec.~\ref{sec:conclusion} summarizes our findings. We use geometrized units $G=c=1$
throughout.

\section{Rotating Black Hole in 4D Einstein--Gauss--Bonnet Gravity}\label{sec:egb}

Einstein--Gauss--Bonnet (EGB) gravity represents one of the simplest extensions of General Relativity by including quadratic curvature corrections in the gravitational action. Although the Gauss--Bonnet invariant is purely topological in four-dimensional spacetime, a consistent regularization procedure allows it to contribute nontrivially to the gravitational dynamics. As a consequence, black hole solutions receive corrections governed by the Gauss--Bonnet coupling parameter $\alpha$, while General Relativity is recovered in the limit $\alpha\rightarrow0$.

The rotating counterpart of the static four-dimensional Einstein--Gauss--Bonnet black hole can be constructed by applying the modified Newman--Janis algorithm. The resulting geometry is stationary, axisymmetric, and asymptotically Kerr-like. In Boyer--Lindquist coordinates $(t,r,\theta,\phi)$, the spacetime metric is given by

\begin{equation}
\begin{aligned}
ds^2 =&
-\left(\frac{\Delta-a^2\sin^2\theta}{\Sigma}\right)dt^2
+\frac{\Sigma}{\Delta}dr^2
+\Sigma d\theta^2  \\
&
-2a\sin^2\theta
\left(
1-\frac{\Delta-a^2\sin^2\theta}{\Sigma}
\right)
dt\,d\phi \\
&
+\sin^2\theta
\left[
\Sigma
+a^2\sin^2\theta
\left(
2-
\frac{\Delta-a^2\sin^2\theta}{\Sigma}
\right)
\right]
d\phi^2 ,
\end{aligned}
\end{equation}

where the metric functions are defined as

\begin{equation}
\Sigma=r^2+a^2\cos^2\theta,
\end{equation}

and

\begin{equation}
\Delta
=
r^2+a^2
+
\frac{r^4}{32\pi\alpha}
\left(
1-
\sqrt{
1+\frac{128\pi\alpha M}{r^3}
}
\right).
\end{equation}

Here, $M$ denotes the black hole mass, $a=J/M$ is the rotational parameter associated with the angular momentum $J$, and $\alpha$ is the Gauss--Bonnet coupling constant. Throughout this work we adopt geometrized units $(G=c=1)$.

The event horizon is determined by the largest positive root of

\begin{equation}
\Delta(r)=0.
\end{equation}

The existence of the horizons depends on both the spin parameter $a$ and the Gauss--Bonnet coupling constant $\alpha$. Depending on these parameters, the equation $\Delta(r)=0$ may admit two positive roots, one degenerate root corresponding to an extremal black hole, or no real positive root.

\begin{figure}[h!]
  \centering
   \includegraphics[height=5cm,width=7cm]{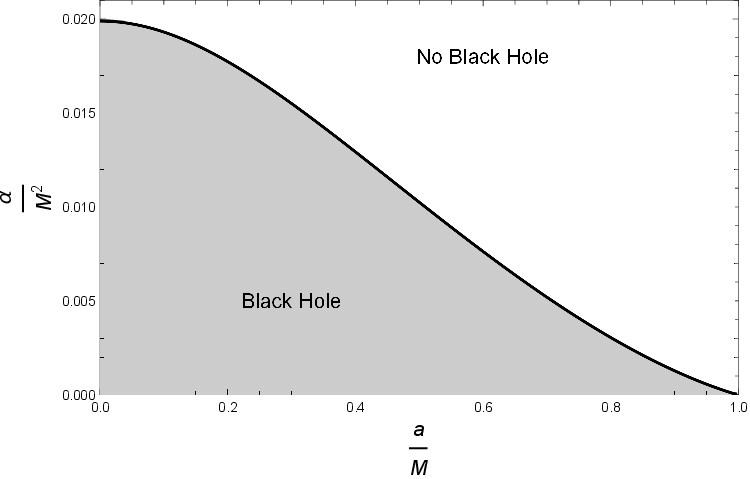}
  \caption{\small{%
  Allowed parameter space for rotating four-dimensional Einstein--Gauss--Bonnet black holes in the $(a/M,\alpha/M^{2})$ plane. The solid curve corresponds to the extremal configuration where the two horizons coincide.}}
  \label{alpha-a}
\end{figure}

Figure~\ref{alpha-a} shows the allowed parameter space in the $(a/M,\alpha/M^{2})$ plane. The solid curve separates black hole solutions from horizonless configurations. The shaded region corresponds to spacetimes with two horizons, whereas the region above the curve does not contain an event horizon and is therefore excluded from our analysis. As the spin increases, the maximum allowed value of $\alpha$ decreases, approaching zero in the Kerr limit ($a/M\rightarrow1$). 
Since our goal is to investigate the maximum rotational energy that can be extracted through the repetitive Penrose process, we concentrate on rapidly rotating black holes with $a/M\lesssim1$. As shown in Fig.~\ref{alpha-a}, this regime corresponds to very small values of the Gauss--Bonnet coupling, $\alpha/M^{2}\ll1$, which will be assumed throughout the analysis.
Since $\alpha/M^{2}\ll1$, the extremal spin parameter can be obtained perturbatively. To leading order in $\alpha$, we find

\begin{equation}
\frac{a_{\rm ext}}{M}\thickapprox 1-32 \pi  \frac{\alpha}{M^2}+1536 \pi ^2 \bigg(\frac{\alpha}{M^2}\bigg)^2-344064 \pi ^3 \bigg(\frac{\alpha}{M^2}\bigg)^3+96862208 \pi ^4 \bigg(\frac{\alpha}{M^2}\bigg)^4 -32879149056 \pi ^5 \bigg(\frac{\alpha}{M^2}\bigg)^5 
\end{equation}

The stationary limit surface is obtained from the condition

$
g_{tt}=0,
$

which does not coincide with the event horizon except at the rotation axis. The region enclosed by the stationary limit surface and the event horizon forms the ergosphere. Since negative-energy states exist inside this region, rotational energy can be extracted through the Penrose process. Therefore, any modification of the ergosphere induced by the Gauss--Bonnet coupling directly influences the efficiency of energy extraction, making this spacetime an appropriate framework for investigating repetitive Penrose processes. 
On the equatorial plane $(\theta=\pi/2)$ and for $\alpha/M^{2}\ll1$, the ergosphere radius is approximated by
\begin{equation}
\frac{r_{ergo}}{M}\thickapprox 2-8 \pi\bigg(\frac{\alpha}{M^2}\bigg)-96 \pi ^2\bigg(\frac{\alpha}{M^2}\bigg)^2
\end{equation}
This approximation shows that the radius of the stationary limit surface decreases as the Gauss--Bonnet coupling increases. Consequently, the ergosphere shrinks relative to that of a Kerr black hole, reducing the region where negative-energy orbits can exist.

The event horizon area is directly related to the irreducible mass, which sets the lower bound on the black hole mass after all possible classical extraction processes have taken place. Following Christodoulou's definition, the irreducible mass is written as

\begin{equation}
M_{\rm irr}
=\sqrt{\frac{A_{\rm H}}{16\pi}}
=\frac{1}{2}\sqrt{r_{+}^{2}+a^{2}}.
\end{equation}

The difference between the total mass and the irreducible mass determines the maximum rotational energy available for extraction,

\begin{equation}
E_{\rm extractable}
=M-M_{\rm irr}
=M-\frac{1}{2}\sqrt{r_{+}^{2}+a^{2}}.
\end{equation}

Unlike the Kerr spacetime, the horizon radius $r_{+}$ is governed by the Gauss--Bonnet coupling through the horizon equation. As a result, both $M_{\rm irr}$ and $E_{\rm extractable}$ inherit an implicit dependence on $\alpha$, indicating that the amount of rotational energy available for extraction is modified by the Gauss--Bonnet correction.

\section{The Repetitive Penrose Process in 4D EGB Gravity}\label{sec:rpp}

We restrict the subsequent analysis to equatorial trajectories, setting
$\theta = \pi/2$. The Penrose process is realized when an incident
particle (particle 0), with rest mass $\mu_0$ and four-momentum
$p_0^\mu$, decays inside the ergoregion into a captured fragment
(particle 1), which falls into the black hole carrying negative energy,
and an escaping fragment (particle 2), which reaches infinity carrying
positive energy. Four-momentum conservation at the decay point then
translates into three independent scalar relations, governing
respectively the energy, the axial angular momentum, and the radial
momentum carried by the three particles involved in the process:
\begin{subequations}
\begin{align}
\hat{E}_0 &= \tilde{\mu}_1 \hat{E}_1 + \tilde{\mu}_2 \hat{E}_2, \\
\hat{p}_{\phi 0} &= \tilde{\mu}_1 \hat{p}_{\phi 1} + \tilde{\mu}_2 \hat{p}_{\phi 2}, \\
\hat{p}_{r 0} &= \tilde{\mu}_1 \hat{p}_{r 1} + \tilde{\mu}_2 \hat{p}_{r 2},
\end{align}
\label{eq:E0P0}
\end{subequations}
where
$\hat{E}_i=E_i/\mu_i$,
$\tilde{\mu}_i=\mu_i/\mu_0$,
$\hat{p}_{\phi i}=p_{\phi i}/(\mu_i M)$,
and
$\hat{p}_{ri}=p_{ri}/\mu_i$
represent the dimensionless energy, the mass ratio, the dimensionless
axial angular momentum, and the dimensionless radial momentum of the
$i$th particle, respectively. These relations hold irrespective of the
detailed form of the background metric; the specific geometry of the
rotating $4D$ EGB spacetime enters only through the normalization
condition and the resulting effective potentials, discussed next.

\subsection{Effective Potential and Turning-Point Condition}

On the equatorial plane, $\theta = \pi/2$, so that $p_\theta = 0$, and the
mass-shell condition $g^{\mu\nu}p_\mu p_\nu = -\mu_i^2$ reduces to
\begin{equation}
g^{tt}\hat{E}_i^2 - 2 g^{t\phi} M \hat{E}_i \hat{p}_{\phi i}
+ g^{\phi\phi} M^2 \hat{p}_{\phi i}^2 + g^{rr}\hat{p}_{ri}^2 = -1 ,
\label{eq:normalization}
\end{equation}
where $g^{tt}$, $g^{t\phi}$, $g^{\phi\phi}$, and $g^{rr}$ denote the inverse
metric components of the rotating $4D$ EGB spacetime evaluated at
$\theta=\pi/2$, and depend on $\hat r$, $\hat a$, and the Gauss--Bonnet
coupling $\hat\alpha$ through the horizon function
$\Delta(\hat r,\hat a,\hat\alpha)$. Solving Eq.~\eqref{eq:normalization} for
the radial momentum gives
\begin{equation}
\hat{p}_{ri}^2 = -\frac{g^{tt}}{g^{rr}}
\left(\hat{E}_i - \hat{V}_i^{+}\right)\left(\hat{E}_i - \hat{V}_i^{-}\right),
\label{eq:pr2}
\end{equation}
with the effective potentials
\begin{equation}
\hat{V}_i^{\pm} = \frac{-g^{t\phi}M\hat{p}_{\phi i}
\pm \sqrt{\left(g^{t\phi}M\hat{p}_{\phi i}\right)^2
- g^{tt}\left(g^{\phi\phi}M^2\hat{p}_{\phi i}^2 + 1\right)}}{g^{tt}} .
\label{eq:Vpm}
\end{equation}
Because $g^{tt}<0$ and $g^{rr}>0$, positivity of $\hat p_{ri}^2$ requires
$\hat E_i \geq \hat V_i^{+}$ or $\hat E_i \leq \hat V_i^{-}$; the requirement
that physical trajectories be future-directed and timelike, $dt/d\tau>0$,
eliminates the lower branch, leaving $\hat V_i^{+}$ as the only branch with
physical meaning.

Maximal energy extraction requires particle~1 to occupy its turning point,
$\hat E_1=\hat V_1^{+}$, the most negative value of $\hat E_1$ compatible
with $\hat E_1\geq \hat V_1^{+}$. Radial momentum conservation,
Eq.~(\ref{eq:E0P0}), then implies $\hat p_{r0}=\hat p_{r2}$. Assuming this
common value to be nonzero and substituting it into Eq.~\eqref{eq:pr2} for
particles 0 and 2, combined with the conservation of energy and angular
momentum, Eq.~(\ref{eq:E0P0}), fixes the energy of the captured particle to
$\hat E_1=(g^{t\phi}/g^{tt})M\hat p_{\phi1}$. Evaluating $dt/d\tau =
-g^{tt}\hat E_i + g^{t\phi}M\hat p_{\phi i}$ at this value gives exactly
$dt/d\tau=0$, in contradiction with the strict inequality $dt/d\tau>0$
required for a physical geodesic. We therefore conclude that
$\hat p_{r0}=\hat p_{r1}=\hat p_{r2}=0$: all three particles must be located
at their respective turning points at the decay point, which fixes the
decay radius $\hat r$ as a free parameter of the process.

Under this triple turning-point condition, and assuming $\hat E_0$,
$\hat p_{\phi 1}$, and the mass ratio $\nu \equiv \mu_2/\mu_1$ are given, the
system of conservation equations, Eq.~\eqref{eq:E0P0}, together with
Eq.~\eqref{eq:normalization} evaluated at $\hat p_{ri}=0$, admits the
following closed-form solution:
\begin{subequations}
\begin{align}
\hat p_{\phi 0} &= \frac{g^{t\phi}\hat E_0 + \sqrt{(g^{t\phi})^2\hat E_0^2
- g^{\phi\phi}(1+g^{tt}\hat E_0^2)}}{Mg^{\phi\phi}} , \\[4pt]
\hat E_1 &= \frac{g^{t\phi}\hat p_{\phi 1}M
- \sqrt{(g^{t\phi})^2\hat p_{\phi 1}^2 M^2
- g^{tt}\left(g^{\phi\phi}\hat p_{\phi 1}^2 M^2 + 1\right)}}{g^{tt}} , \\[4pt]
\tilde\mu_1 &= \frac{\hat E_0\hat E_1 g^{tt} - \hat E_1 g^{t\phi}M\hat
p_{\phi 0} - \hat E_0 g^{t\phi}M\hat p_{\phi 1}
+ g^{\phi\phi}M^2\hat p_{\phi 0}\hat p_{\phi 1} + \sqrt{D}}
{\hat E_1^2 g^{tt} - 2\hat E_1 g^{t\phi}M\hat p_{\phi 1}
+ g^{\phi\phi}M^2\hat p_{\phi 1}^2 + \nu^2} , \\[4pt]
\hat E_2 &= \frac{\hat E_0}{\tilde\mu_2} - \frac{\hat E_1}{\nu} , \qquad
\hat p_{\phi 2} = \frac{\hat p_{\phi 0}}{\tilde\mu_2} - \frac{\hat
p_{\phi 1}}{\nu} ,
\end{align}
\label{eq:analytic-solution}
\end{subequations}
with the discriminant
\begin{align}
D ={}& -g^{tt}g^{\phi\phi}M^2\hat E_1^2\hat p_{\phi 0}^2
+ (g^{t\phi})^2M^2\hat E_1^2\hat p_{\phi 0}^2
- g^{tt}g^{\phi\phi}M^2\hat E_0^2\hat p_{\phi 1}^2
+ (g^{t\phi})^2M^2\hat E_0^2\hat p_{\phi 1}^2 \notag \\
&+ 2g^{tt}g^{\phi\phi}M^2\hat E_0\hat E_1\hat p_{\phi 0}\hat p_{\phi 1}
- 2(g^{t\phi})^2M^2\hat E_0\hat E_1\hat p_{\phi 0}\hat p_{\phi 1}
- g^{tt}\hat E_0^2\nu^2 + 2g^{t\phi}M\hat E_0\hat p_{\phi 0}\nu^2 \notag \\
&- g^{\phi\phi}M^2\hat p_{\phi 0}^2\nu^2 .
\label{eq:D-discriminant}
\end{align}
This solution reduces to the corresponding expressions for the Kerr
spacetime once $g^{tt}$, $g^{t\phi}$, and $g^{\phi\phi}$ are evaluated at
$\hat\alpha=0$, and its functional form is generic to any stationary,
axisymmetric metric admitting a $(t,\phi)$ Killing pair, with the specific
geometry of the $4D$ EGB black hole entering only through the metric
functions themselves.

\subsection{Mass, Spin, and Effective Coupling Evolution in the Iterative Process}\label{sec:analytic-solution}

Following each Penrose decay, the black hole absorbs the negative-energy
fragment, leading to a simultaneous change in its mass and angular
momentum. If $(M_{n-1},L_{n-1})$ denote the black-hole parameters
immediately before the $n$th decay, conservation of energy and angular
momentum gives

\begin{align}\label{eq:update}
M_n &= M_{n-1} + \mu_0\,\tilde{\mu}_{1,n-1}\,\hat E_{1,n-1}, \\
L_n &= L_{n-1} + M_{n-1}\,\mu_{1,n-1}\,\hat p_{\phi1},
\end{align}

where the subscript $(n-1)$ emphasizes that the conserved quantities are
evaluated using the black-hole parameters from the previous iteration.
Since the captured particle satisfies $\hat E_{1,n-1}<0$ and
$\hat p_{\phi1}<0$, both the mass and the angular momentum decrease
monotonically, $M_n<M_{n-1}$ and $L_n<L_{n-1}$.

The corresponding evolution of the dimensionless spin parameter,
$\hat a_n=L_n/M_n^2$, is governed by
\begin{equation}
\Delta\hat a_{n-1} = \frac{L_n}{M_n^2} - \frac{L_{n-1}}{M_{n-1}^2}.
\label{eq:spinupdate}
\end{equation}
Unlike the mass and angular momentum, whose variations follow directly
from the conservation laws, the evolution of $\hat a$ is intrinsically
nonlinear, since the angular momentum enters linearly whereas the mass
enters quadratically through the denominator.

Unlike the black-hole mass and angular momentum, the Gauss--Bonnet
coupling constant $\alpha$ is not a quantity carried by the infalling
particles; it is a fixed parameter of the underlying gravitational
action and therefore remains unchanged during the repetitive Penrose
process. However, the spacetime is naturally described in terms of the
dimensionless coupling
$
\hat{\alpha} = \frac{\alpha}{M^2},
$
which must be updated after every iteration according to
\begin{equation}
\hat{\alpha}_n = \frac{\alpha}{M_n^2}.
\label{eq:alphaupdate}
\end{equation}
Because the black-hole mass decreases monotonically, it follows
immediately that
$
\hat{\alpha}_n > \hat{\alpha}_{n-1} .
$
Consequently, each Penrose event modifies not only the mass and spin of
the black hole but also the effective strength of the Gauss--Bonnet
correction, through the increase of the dimensionless coupling
$\hat\alpha$. The spacetime geometry, including the horizon radius and
all quantities derived from it, must therefore be recalculated at every
iteration using the fully updated set of parameters
$(M_n,\hat a_n,\hat\alpha_n)$.

The cumulative energy extracted after the first $n$ Penrose events is
given by the total decrease in the black-hole mass,
\begin{equation}
E_{\mathrm{extracted},n} = M_0-M_n,
\label{eq:Eextracted}
\end{equation}
where $M_0$ denotes the initial black-hole mass before the onset of the
repetitive Penrose process. Because the irreducible mass $M_{\rm irr}$ is
itself a nonlinear function of $(M_n,\hat a_n,\hat\alpha_n)$, the
recursive updates in Eqs.~(\ref{eq:update})--(\ref{eq:alphaupdate})
directly control how much of the extracted energy in
Eq.~(\ref{eq:Eextracted}) is drawn from the reservoir of extractable
energy, as opposed to being irreversibly stored in the horizon area --
a connection made explicit in the following subsection.
To evaluate the iterative energy extraction process, we introduce two primary metrics. The energy return on investment (EROI), $\xi_n$, quantifies the net extracted energy relative to the total energy supplied by incident particles over $n$ iterations:
\begin{equation}
\xi_n = \frac{E_{\text{extracted},n}}{n E_0}.
\end{equation}
Concurrently, the energy utilization efficiency (EUE), $\Xi_n$, measures the harvested energy against the actual depletion of the black hole's extractable reservoir, defined by the difference between the initial and current extractable limits:
\begin{equation}
\Xi_n = \frac{E_{\text{extracted},n}}{E_{\text{extractable},0} - E_{\text{extractable},n}}.
\end{equation}
Sustaining this repetitive mechanism fundamentally requires a positive mass deficit, ensuring the original particle's mass exceeds the combined mass of the resulting decay products:
\begin{equation}
\mu_0 - \mu_1 - \mu_2 > 0.
\end{equation}
While this mass condition ensures the kinematic feasibility of the decay, the global sustainability of the extraction sequence is ultimately dictated by the black hole's rotational state.
\subsection{Critical Spin Thresholds and Termination of the Repetitive Process}

The recursive evolution described above cannot be carried out
indefinitely. With each Penrose event the black-hole mass and angular
momentum decrease, while the dimensionless Gauss--Bonnet coupling
$\hat\alpha_n=\alpha/M_n^2$ increases correspondingly, so that the
background geometry changes continuously over the course of the
sequence. This continuous deformation modifies the effective potentials
governing the three particles, and the process terminates once no
configuration compatible with the common turning-point condition can be
found.

The point at which this occurs is determined independently for each of
the three particles participating in the decay. The incident particle
must be able to reach the splitting point from infinity, the captured
fragment must cross the event horizon, and the escaping fragment must
remain unbound; in each case, the critical orbit corresponds to the
configuration in which the turning point coincides with the maximum of
the respective effective potential,
\begin{equation}
\hat V_i^{+}(\hat r)=\hat E_i , \qquad
\frac{d\hat V_i^{+}}{d\hat r}=0 , \qquad i=0,1,2 .
\label{eq:marginal}
\end{equation}
Solving this condition separately for each particle yields three
independent critical spin parameters, $\hat a_{\min,0}$,
$\hat a_{\min,1}$, and $\hat a_{\min,2}$. Since the decay requires all
three particles to satisfy their respective orbital constraints
simultaneously, the repetitive process can proceed only while the
black-hole spin remains above the largest of the three.

For the incident particle, released from rest at infinity so that
$\hat E_0=1$, combining Eq.~\eqref{eq:marginal} with the analytic
solution of Sec.~\ref{sec:analytic-solution} gives
\begin{equation}
\hat{a}_{\min,0} =
\sqrt{\,
\hat r^{2}+4\hat r+\frac{96\pi\hat\alpha\,(2\hat r+3)\hat r}{\hat r^{3}-128\pi\hat\alpha}
-\frac{32\pi\hat\alpha}{\hat r^{2}}
-\frac{4 (\hat r^{3} - 32\pi\hat\alpha) \sqrt{64\pi\hat\alpha+\hat r^{3}}}{\hat r^{3}-128\pi\hat\alpha}
\,}\; .
\label{eq:amin0-EGB}
\end{equation}
In the limit $\hat\alpha\to0$, Eq.~\eqref{eq:amin0-EGB} reduces to the
standard Kerr expression, $\hat a_{\min,0}=2\sqrt{\hat r}-\hat r$,
confirming the consistency of the result.

The bound associated with the captured particle follows from the
requirement that the decay take place outside the outer horizon,
$\hat\Delta(\hat r,\hat a,\hat\alpha)=0$, and is given by
\begin{equation}
\hat a_{\min,1} =\sqrt{2\hat{r}-\hat{r}^2-\frac{64 \pi \hat{\alpha}}{\hat{r}^2}},
\label{eq:amin1-EGB}
\end{equation}
which reduces to $\hat a_{\min,1}=\sqrt{\hat r(2-\hat r)}$ for
$\hat\alpha=0$. In contrast to Eq.~\eqref{eq:amin0-EGB}, the
Gauss--Bonnet coupling enters here only as an additive term beneath the
square root, a consequence of the fact that this bound is determined
directly by the location of the horizon rather than by the structure of
an effective potential.

The bound associated with the escaping particle applies when the
incident particle carries $\hat E_0>1$, in which case the limiting orbit
is the equatorial corotating photon orbit, determined by
$(\hat r\hat\Delta'-4\hat\Delta)^2=16\hat a^2\hat\Delta$. Evaluating this
condition for the horizon function of the rotating $4D$ EGB spacetime
gives
\begin{equation}
\hat a_{\min,2} =
\frac{3\hat r^{3}-\hat r^{4}-192\pi\hat\alpha}
{2\hat r\sqrt{\hat r^{3}-128\pi\hat\alpha}} ,
\label{eq:amin2-EGB}
\end{equation}
which reduces to $\hat a_{\min,2}=(3-\hat r)\sqrt{\hat r}/2$ in the same
limit.

\begin{figure}[h!]
  \centering
  \begin{subfigure}[b]{0.48\textwidth}
    \centering
    \includegraphics[height=5cm,width=7cm]{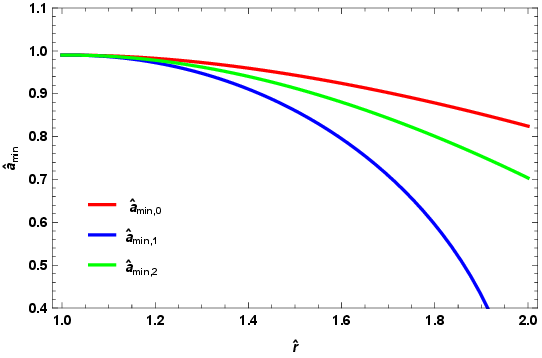}
    \caption{}
    \label{ami012GB}
  \end{subfigure}
  \hfill
  \begin{subfigure}[b]{0.48\textwidth}
    \centering
    \includegraphics[height=5cm,width=7cm]{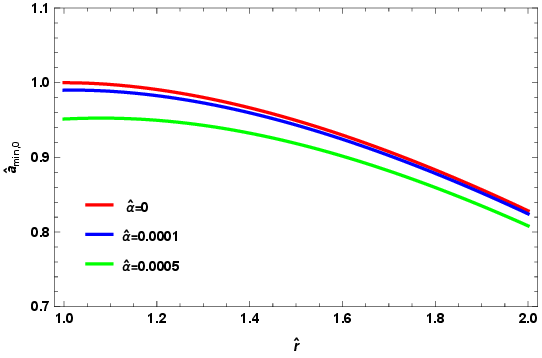}
    \caption{}
    \label{ami0alpha}
  \end{subfigure}
  \caption{\small{%
  (\subref{ami012GB}) The three critical spin thresholds $\hat a_{\min,0}$,
 $\hat a_{\min,1}$, and $\hat a_{\min,2}$ as functions of the dimensionless
decay radius $\hat r$, for fixed Gauss--Bonnet coupling
 $\hat\alpha=0.0001$ and $\hat E_0=1$. (\subref{ami0alpha}) The critical spin threshold
 $\hat a_{\min,0}$ as a function of $\hat r$ for three representative values of
 $\hat\alpha=0,\,0.0001,\,0.0005$.}}
  \label{fig:amin}
\end{figure}

The physically relevant threshold is the largest of the three bounds,
\begin{equation}
\hat a_{\min} = \max\{\hat a_{\min,0},\,\hat a_{\min,1},\,\hat a_{\min,2}\} .
\label{eq:amin-max}
\end{equation}
Figure~\ref{ami012GB} shows the three critical spin values for
$\hat\alpha=10^{-4}$. Throughout the interval $1\lesssim\hat r\lesssim2$,
the ordering $\hat a_{\min,1}<\hat a_{\min,2}<\hat a_{\min,0}$ is
satisfied, so that the incident particle -- rather than the captured or
escaping fragment -- determines the termination of the repetitive
process. This ordering was verified to hold throughout the parameter
space considered in this work, indicating that, unlike the
electromagnetic coupling in the charged Kerr--Newman case \cite{Aliporcharged2026}, the
Gauss--Bonnet coupling does not alter which particle governs the
stopping condition; it modifies only the magnitude of the bound itself.

Figure~\ref{ami0alpha} shows $\hat a_{\min,0}$ as a function of
$\hat r$ for several values of $\hat\alpha$. At fixed decay radius, the
critical spin decreases monotonically with increasing $\hat\alpha$.
Taken in isolation, this trend would suggest that a larger
Gauss--Bonnet coupling permits the repetitive process to proceed to
lower spins before termination. However, the initial spin $\hat a_0$ of
the extremal black hole is itself a function of $\hat\alpha$, and the
interplay between $\hat a_0$ and $\hat a_{\min,0}$ determines the size
of the interval available to the iteration, as quantified in
Sec.~\ref{sec:numerical-results}.

\newpage
\section{ Numerical Analysis of the Repetitive Penrose Process in the Rotating 4D EGB Spacetime}\label{sec:numerical-results}
In this section, we carry out a detailed numerical study of the
repetitive Penrose process in the rotating $4D$ EGB spacetime,
following the same iterative scheme developed in
Sec.~\ref{sec:analytic-solution}. As in previous treatments of the
repetitive Penrose process, we set the incident energy to
$\hat E_0=1$; a particle released from rest at infinity is known to
yield a higher energy return on investment (EROI) than one launched with
$\hat E_0>1$, and we adopt this choice throughout to maximize
the extraction efficiency Ref.~\cite{Ruffini2025PRR}.

For the incident particle and the decay configuration, we take
$\hat p_{\phi 1}=-19.434$, a mass ratio $\nu=\mu_2/\mu_1=0.78345$, and a
rest mass $\mu_0=10^{-2}M_0$ for the incident particle, matching the
choice used in Ref.~\cite{Ruffini2025PRR} for the Kerr case; this allows
a direct comparison between the Kerr and EGB results once
$\hat\alpha\to0$. The decay is taken to occur at the fixed dimensionless radius
$\hat r=1.2$; the corresponding results for $\hat r=1.5$ are reported
in Appendix~\ref{app:r15} for completeness.

With these parameters fixed, the analytic expressions of
Eqs.~(\ref{eq:analytic-solution})--(\ref{eq:D-discriminant}) determine the outcome of each
decay, after which the black-hole mass $M_n$, angular momentum $L_n$,
and dimensionless Gauss--Bonnet coupling $\hat\alpha_n$ are updated
according to Eqs.~(\ref{eq:update})--(\ref{eq:alphaupdate}). Repeating
this procedure step by step generates the full iterative sequence,
whose results we present in the tables and figures below.
 In particular, we examine how the value of the Gauss--Bonnet
coupling $\hat\alpha$ affects the number of decays the black hole can
sustain, the resulting extraction efficiency, and the final black-hole
parameters at the end of the sequence.

 \begin{table}[htbp]
    \centering
    \setlength{\arrayrulewidth}{0.5pt}
    \renewcommand{\arraystretch}{1.2}
    \scriptsize
    \setlength{\tabcolsep}{3pt}
\resizebox{\textwidth}{!}{%
\begin{tabular}{cccccccccccccc}
\toprule
 $\hat{\alpha}_0$ & $n$ & $M/M_0$ & $\hat{a}_n$ & $\hat{\alpha}_n$ & $E_{\text{extractable}}/M_0$ & $E_{\text{extracted}}/M_0$ & $M_\text{irr}/M_0$ & $\tilde{\mu}_{1,n}$ & $\hat{E}_{1,n}$ & $\hat{p}_{\phi0,n}$ & $\hat{a}_{\min,0,n}$ & $\xi_n$ & $\Xi_n$  \\
\midrule

\multirow{10}{*}{0} & 0 & 1 & 1 & 0 & 0.292893 & 0 & 0.707107 & 0.0209006 & -6.93558 & 2.1127 & 0.99089 & 0 & 0 \\
 & 1 & 0.99855 & 0.998832 & 0 & 0.275611 & 0.00144958 & 0.72294 & 0.0208126 & -6.96513 & 2.12126 & 0.99089 & 0.144958 & 0.0838746 \\
 & 2 & 0.997101 & 0.997676 & 0 & 0.268419 & 0.0028992 & 0.728682 & 0.0207235 & -6.99529 & 2.13006 & 0.99089 & 0.14496 & 0.11846 \\
 & 3 & 0.995651 & 0.996532 & 0 & 0.262915 & 0.00434887 & 0.732736 & 0.0206333 & -7.0261 & 2.13914 & 0.99089 & 0.144962 & 0.145066 \\
 & 4 & 0.994201 & 0.995402 & 0 & 0.258294 & 0.00579858 & 0.735907 & 0.0205417 & -7.05764 & 2.14852 & 0.99089 & 0.144965 & 0.167593 \\
 & 5 & 0.992752 & 0.994284 & 0 & 0.254245 & 0.00724834 & 0.738506 & 0.0204488 & -7.08996 & 2.15822 & 0.99089 & 0.144967 & 0.187549 \\
 & 6 & 0.991302 & 0.99318 & 0 & 0.250608 & 0.00869815 & 0.740694 & 0.0203542 & -7.12314 & 2.16827 & 0.99089 & 0.144969 & 0.205702 \\
 & 7 & 0.989852 & 0.992089 & 0 & 0.247286 & 0.010148 & 0.742566 & 0.0202578 & -7.15727 & 2.17872 & 0.99089 & 0.144971 & 0.22251 \\
 & 8 & 0.988402 & 0.991013 & 0 & 0.244218 & 0.0115979 & 0.744184 & 0.0201593 & -7.19248 & 2.18961 & 0.99089 & 0.144974 & 0.238271 \\
 & \textcolor{red}{9} & \textcolor{red}{0.986952} & \textcolor{red}{0.989952} & \textcolor{red}{0} & \textcolor{red}{0.24136} & \textcolor{red}{0.0130479} & \textcolor{red}{0.745592} & \textcolor{red}{0.0200585} & \textcolor{red}{-7.22889} & \textcolor{red}{2.20099} & \textcolor{red}{0.99089} & \textcolor{red}{0.144976} & \textcolor{red}{0.253193} \\
\midrule

\multirow{6}{*}{0.0003} & 0 & 1 & 0.970969 & 0.0003 & 0.283057 & 0 & 0.716943 & 0.0196619 & -7.08811 & 2.17863 & 0.966225 & 0 & 0 \\
 & 1 & 0.998606 & 0.969849 & 0.0003 & 0.268021 & 0.00139365 & 0.730585 & 0.0195617 & -7.12465 & 2.18996 & 0.966225 & 0.139365 & 0.0926895 \\
 & 2 & 0.997213 & 0.968745 & 0.000300838 & 0.261669 & 0.00278736 & 0.735543 & 0.0194631 & -7.16018 & 2.20115 & 0.966156 & 0.139368 & 0.130327 \\
 & 3 & 0.995819 & 0.967654 & 0.000302522 & 0.257058 & 0.00418095 & 0.738761 & 0.0193663 & -7.19455 & 2.21216 & 0.966018 & 0.139365 & 0.160816 \\
 & 4 & 0.994426 & 0.966578 & 0.000305068 & 0.253406 & 0.00557427 & 0.74102 & 0.0192717 & -7.22759 & 2.22293 & 0.965808 & 0.139357 & 0.187998 \\
 & \textcolor{red}{5} & \textcolor{red}{0.993033} & \textcolor{red}{0.965514} & \textcolor{red}{0.000308497} & \textcolor{red}{0.250407} & \textcolor{red}{0.00696715} & \textcolor{red}{0.742625} & \textcolor{red}{0.0191798} & \textcolor{red}{-7.25908} & \textcolor{red}{2.23336} & \textcolor{red}{0.965525} & \textcolor{red}{0.139343} & \textcolor{red}{0.213393} \\
\midrule

\multirow{3}{*}{0.0006} & 0 & 1 & 0.943275 & 0.0006 & 0.269958 & 0 & 0.730042 & 0.0184799 & -7.23575 & 2.25091 & 0.941426 & 0 & 0 \\
 & 1 & 0.998663 & 0.942202 & 0.0006 & 0.260736 & 0.00133716 & 0.737926 & 0.0183581 & -7.28403 & 2.26692 & 0.941426 & 0.133716 & 0.14501 \\
 & \textcolor{red}{2} & \textcolor{red}{0.997326} & \textcolor{red}{0.941148} & \textcolor{red}{0.000601608} & \textcolor{red}{0.255555} & \textcolor{red}{0.00267437} & \textcolor{red}{0.74177} & \textcolor{red}{0.0182413} & \textcolor{red}{-7.32929} & \textcolor{red}{2.28229} & \textcolor{red}{0.941293} & \textcolor{red}{0.133718} & \textcolor{red}{0.18569} \\
\midrule

0.0012 & \multicolumn{13}{c}{\textit{Repetitive Penrose process does not occur}} \\

\bottomrule
\end{tabular}
}
\caption{Repetitive Penrose process for neutral particles at $\hat{r}=1.2$ for various initial $\hat{\alpha}_0$ in rotating black holes in 4D Einstein-Gauss-Bonnet gravity. The rows shown in red correspond to the first inadmissible iteration ($n=n_f+1$), for which $\hat a_n<\hat a_{\min,0,n}$. Fixed parameters: $\hat{E}_0=1$, $\hat{r}=1.2$,
$\nu=0.78345$, $\hat{p}_{\phi1}=-19.434$, $\mu_0=0.01M_0$, $M_0=1$.}
\label{tab:combined_alpha_r1.2}
\end{table}

\begin{figure}[htbp]
  \centering
  \begin{subfigure}[b]{0.48\textwidth}
    \centering
    \includegraphics[height=5cm,width=\textwidth]{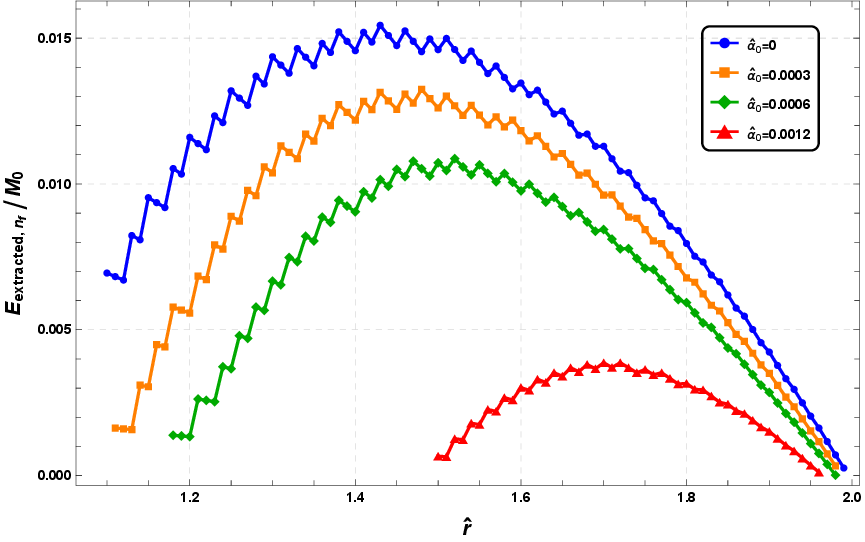}
    \caption{}
    \label{Eextracted}
  \end{subfigure}
  \hfill
  \begin{subfigure}[b]{0.48\textwidth}
    \centering
    \includegraphics[height=5cm,width=\textwidth]{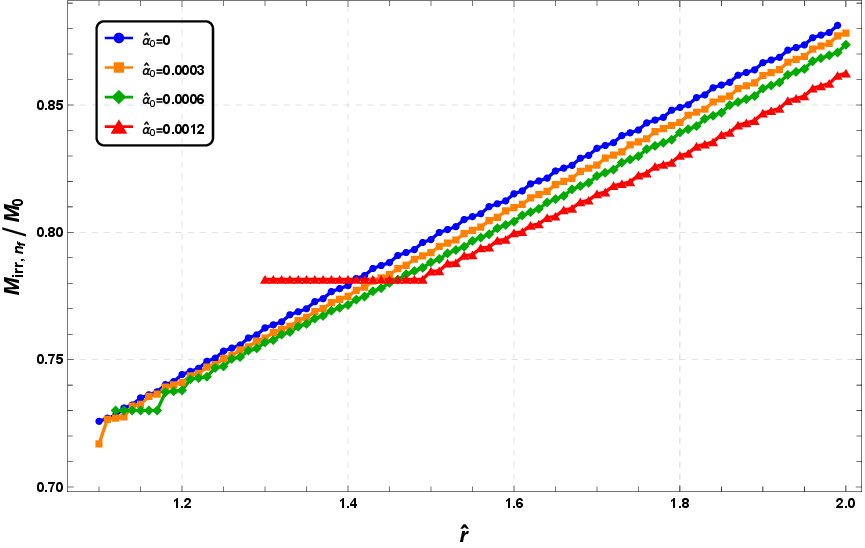}
    \caption{}
    \label{Mirr}
  \end{subfigure}
  \caption{\small{%
  (\subref{Eextracted}) Final extracted energy $E_{\text{extracted},n_f}/M_0$ and (\subref{Mirr}) final irreducible mass $M_{\text{irr}, n_f}/M_0$ of the repetitive Penrose process as functions of the dimensionless decay radius $\hat{r}$ for different values of the Gauss-Bonnet coupling parameter $\hat{\alpha}_0$.}}
  \label{fig:energy_irr}
\end{figure}

Figure~\ref{Eextracted} demonstrates that increasing the Gauss--Bonnet 
coupling $\hat{\alpha}_0$ shifts the onset of energy extraction to 
larger radii, near the ergosphere. Across all coupling strengths, 
the intermediate radial regime consistently yields the maximum total 
extracted energy. Concurrently, an examination of 
Fig.~\ref{Mirr} reveals that while the irreducible mass 
 $M_{\text{irr},n_f}/M_0$ invariably increases once the decay sequence 
is initiated, its growth is systematically suppressed in the EGB 
spacetime compared to the Kerr baseline as $\hat{\alpha}_0$ increases.

\begin{figure}[htbp]
  \centering
  \begin{subfigure}[b]{0.48\textwidth}
    \centering
    \includegraphics[height=5cm,width=\textwidth]{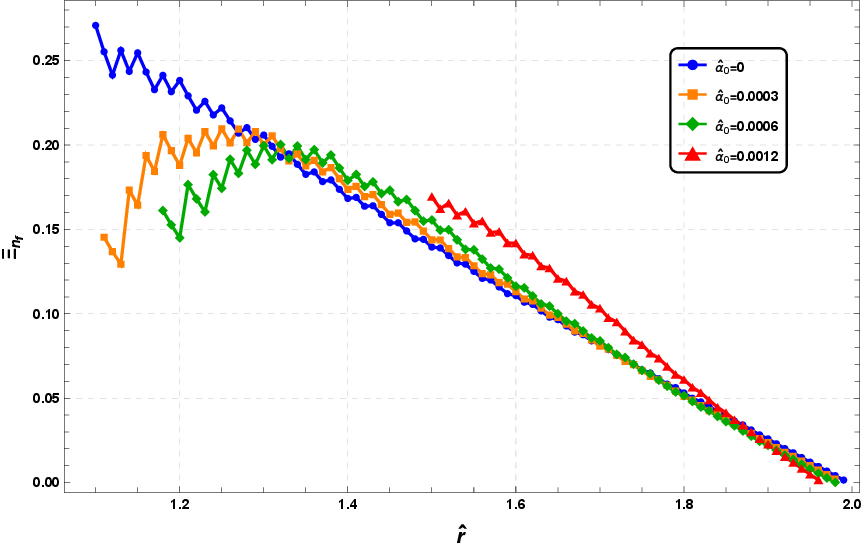} 
    \caption{}
    \label{fig:EUE}
  \end{subfigure}
  \hfill
  \begin{subfigure}[b]{0.48\textwidth}
    \centering
    \includegraphics[height=5cm,width=\textwidth]{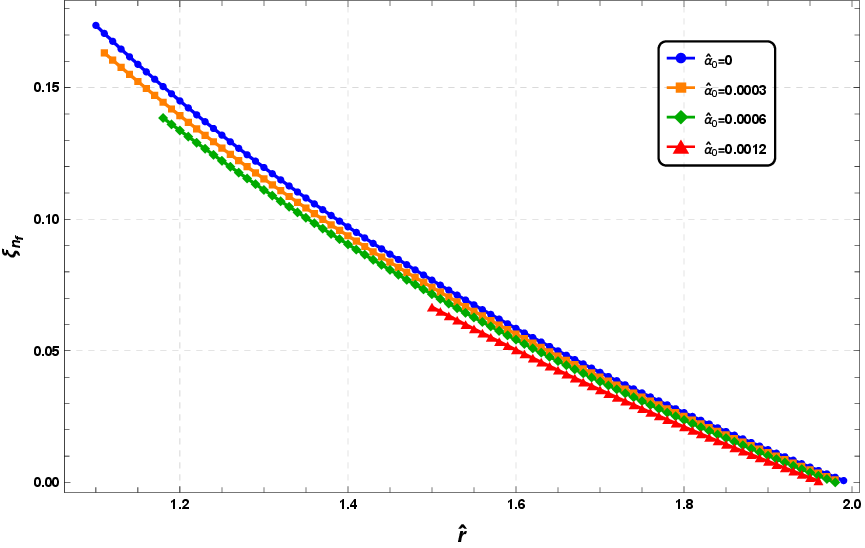}
    \caption{}
    \label{fig:EROI}
  \end{subfigure}
  \caption{\small{%
  (\subref{fig:EUE}) Final energy utilization efficiency (EUE), $\Xi$, and (\subref{fig:EROI}) final energy return on investment (EROI), $\xi$, of the repetitive Penrose process as functions of the dimensionless decay radius $\hat{r}$ for different values of the Gauss-Bonnet coupling parameter $\hat{\alpha}_0$.}}
  \label{fig:eue_eroi}
\end{figure}

\begin{figure}[htbp]
  \centering
 
  \begin{subfigure}[b]{0.48\textwidth}
    \centering
    \includegraphics[height=5cm,width=\textwidth]{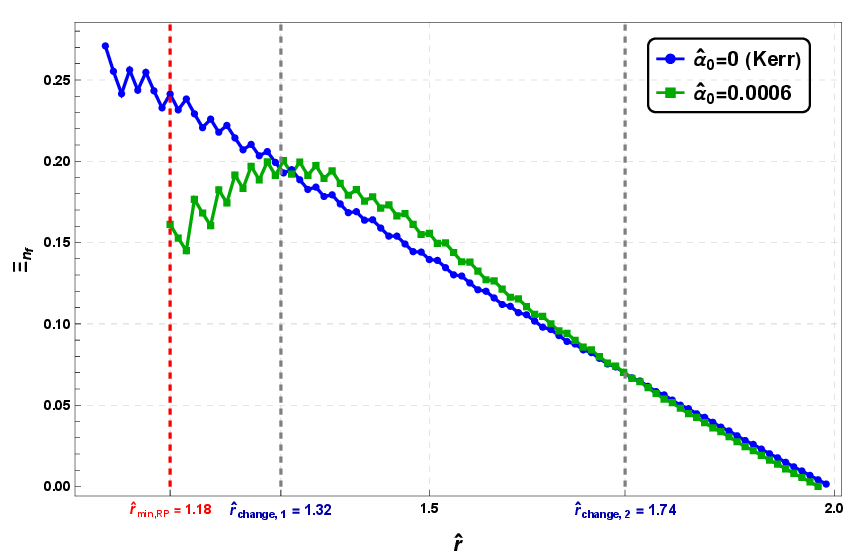}
    \caption{}
    \label{change1,2}
  \end{subfigure}
  \hfill
  \begin{subfigure}[b]{0.48\textwidth}
    \centering
    \includegraphics[height=5cm,width=\textwidth]{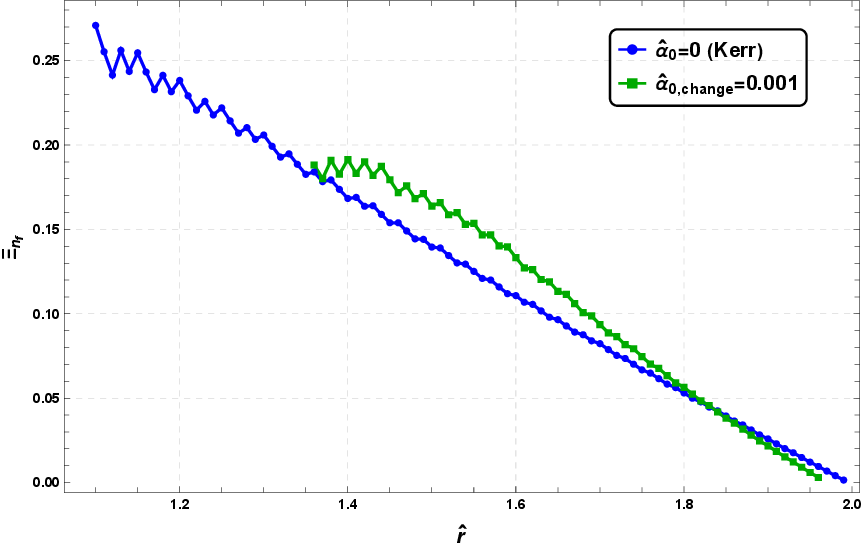}
    \caption{}
    \label{alphachange}
  \end{subfigure}
  
  \vspace{0.4cm}

  \begin{subfigure}[b]{0.48\textwidth}
    \centering
    \includegraphics[height=5cm,width=\textwidth]{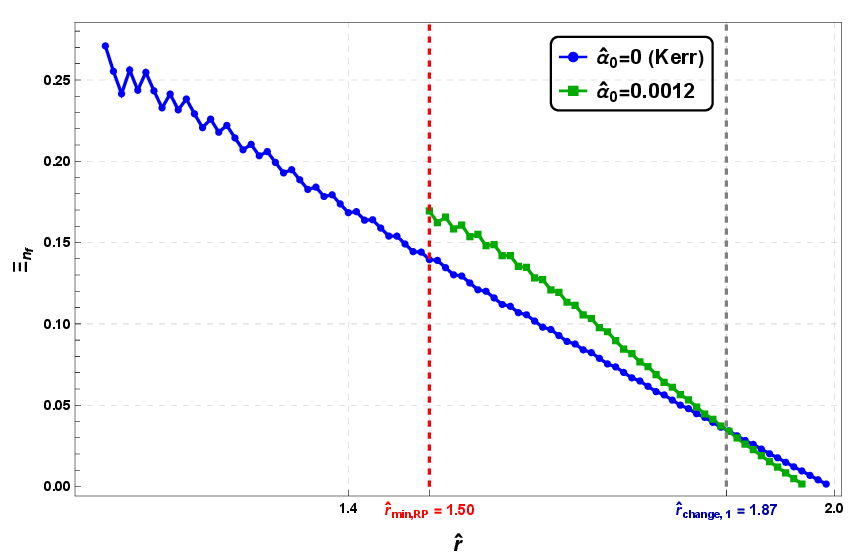}
    \caption{}
    \label{change1}
  \end{subfigure}
  
  \caption{\small{%
 (\subref{change1,2}) Final efficiency $\Xi_{nf}$ as a function of the decay radius $\hat{r}$ for $\hat{\alpha}_0 < \hat{\alpha}_{0,change}$ (e.g., $0.0006$), exhibiting a four-region structure. The process is null for EGB below $\hat{r}_{\text{min,RP}}$ (while remaining active for Kerr), followed by a suppressed regime ($\Xi_{\text{GB}} < \Xi_{\text{Kerr}}$), a bounded enhanced regime ($\hat{r}_{\text{change},1} < \hat{r} < \hat{r}_{\text{change},2}$), and a second suppressed regime for $\hat{r} > \hat{r}_{\text{change},2}$. 
 (\subref{alphachange}) The critical transition regime at $\hat{\alpha}_0 = \hat{\alpha}_{0,change} \approx 0.001$, where the two change radii merge into a single tangency point, marking the threshold for the topological transition. 
  (\subref{change1}) Final efficiency for $\hat{\alpha}_0 > \hat{\alpha}_{0,change}$ (e.g., $0.0012$), displaying a three-region structure with an inverted initial behavior ($\Xi_{\text{GB}} > \Xi_{\text{Kerr}}$ immediately after the shifted $\hat{r}_{\text{min,RP}}$), before crossing below the Kerr efficiency at the single remaining threshold $\hat{r}_{\text{change},1}$.
 }}
  \label{fig:change}
\end{figure}

Conversely, the energy return on investment ($\xi$, 
Fig.~\ref{fig:EROI}) exhibits a monotonic decrease with increasing 
 $\hat{\alpha}_0$, with the optimal return shifting toward larger decay 
radii. The energy utilization efficiency ($\Xi$, 
Fig.~\ref{fig:EUE}), however, displays a distinctly non-monotonic 
topology that diverges from the behavior of the other metrics. This 
structural variation implies that the comparative performance between 
the EGB and Kerr black holes is not uniform, but instead divides the 
parameter space into distinct radial regimes depending on 
 $\hat{\alpha}_0$. This necessitates a more granular investigation of 
the efficiency profiles, which we present in Fig.~\ref{fig:change}.

As detailed in Fig.~\ref{change1,2}, for a sub-critical coupling such as 
 $\hat{\alpha}_0 = 0.0006$, the parameter space divides into four 
distinct regimes relative to the Kerr efficiency. Below the minimum 
permissible radius, $\hat{r} < \hat{r}_{\text{min,RP}}$, the repetitive 
Penrose process is kinematically forbidden for the EGB black hole, 
though it remains active for the Kerr metric. In the subsequent regime, 
 $\hat{r}_{\text{min,RP}} < \hat{r} < \hat{r}_{\text{change},1}$, the 
EGB efficiency falls below the Kerr baseline. This is followed by a 
bounded window, $\hat{r}_{\text{change},1} < \hat{r} < 
\hat{r}_{\text{change},2}$, where the EGB efficiency surpasses that 
of the Kerr black hole. For decay radii exceeding 
 $\hat{r}_{\text{change},2}$, the Kerr efficiency once again becomes 
dominant. 

As $\hat{\alpha}_0$ is increased further, $\hat{r}_{\text{min,RP}}$ 
shifts progressively outward. This trend culminates at the critical 
transition value, $\hat{\alpha}_0 = \hat{\alpha}_{0,change}$, where 
the four-region structure collapses into a three-region configuration. 
In this regime, immediately above the shifted minimum radius 
($\hat{r}_{\text{min,RP}} < \hat{r} < \hat{r}_{\text{change},1}$), 
the EGB efficiency initially exceeds the Kerr limit; however, beyond 
the single remaining crossover point ($\hat{r} > 
\hat{r}_{\text{change},1}$), the Kerr efficiency regains dominance.

\begin{table}[htbp]
    \centering
    \caption{Classification of the final efficiency regimes for the repetitive Penrose process in 4D EGB spacetime, categorized by the Gauss--Bonnet coupling parameter $\hat{\alpha}_0$ and the dimensionless decay radius $\hat{r}$. The critical transition value is denoted as $\hat{\alpha}_{0,change} \approx 0.001$.}
    \label{tab:efficiency_regimes}
    \renewcommand{\arraystretch}{1.4}
    \small
    \begin{tabular}{@{} c l c l @{}}
    \toprule
    \textbf{Coupling Regime} & \textbf{Range of $\hat{r}$} & \textbf{Efficiency Relation} & \textbf{Physical Interpretation} \\
    \midrule
    
    \multirow{4}{*}{$\hat{\alpha}_0 < \hat{\alpha}_{0,change}$} 
    & $\hat{r} < \hat{r}_{\text{min,RP}}$ & $\Xi_{\text{GB}} = 0$ & \textit{Null for EGB} (Active for Kerr) \\
    & $\hat{r}_{\text{min,RP}} < \hat{r} < \hat{r}_{\text{change},1}$ & $\Xi_{\text{GB}} < \Xi_{\text{Kerr}}$ & Suppressed regime \\
    & $\hat{r}_{\text{change},1} < \hat{r} < \hat{r}_{\text{change},2}$ & \textcolor{red}{$\Xi_{\text{GB}} > \Xi_{\text{Kerr}}$} & \textcolor{red}{Enhanced regime} (Efficiency island) \\
    & $\hat{r} > \hat{r}_{\text{change},2}$ & $\Xi_{\text{GB}} < \Xi_{\text{Kerr}}$ & Suppressed regime \\
    
    \midrule
    
    \multirow{3}{*}{$\hat{\alpha}_0 \ge \hat{\alpha}_{0,change}$} 
    & $\hat{r} < \hat{r}_{\text{min,RP}}$ & $\Xi_{\text{GB}} = 0$ & \textit{Null for EGB} (Active for Kerr) \\
    & $\hat{r}_{\text{min,RP}} < \hat{r} < \hat{r}_{\text{change},1}$ & \textcolor{red}{$\Xi_{\text{GB}} > \Xi_{\text{Kerr}}$} & \textcolor{red}{Enhanced regime} (Inverted behavior) \\
    & $\hat{r} > \hat{r}_{\text{change},1}$ & $\Xi_{\text{GB}} < \Xi_{\text{Kerr}}$ & Suppressed regime \\
    
    \bottomrule
    \end{tabular}
\end{table}

\section{Conclusion}
\label{sec:conclusion}

We have studied the repetitive Penrose process for neutral particles in a rotating
four-dimensional Einstein--Gauss--Bonnet black hole, treating every decay
self-consistently through the triple turning-point condition and updating the mass,
angular momentum, and irreducible mass at each step. The analytic solution of
Sec.~\ref{sec:rpp} reduces to the Kerr result of Ref.~\cite{Ruffini2025PRR} as
$\hat\alpha\to0$, which served both as a consistency check and as the baseline for all
comparisons.

What sets this case apart from the earlier extensions of the process is the status of
the coupling. In the charged Kerr--Newman geometry~\cite{Aliporcharged2026} the
electromagnetic interaction enters through a quantity carried by the fragments, while in
the Kerr--de~Sitter~\cite{WangZeng2025} and accelerating Kerr~\cite{ZengWang2026} cases
the additional parameter is an external constant fixed once and for all. The
Gauss--Bonnet coupling $\alpha$ belongs to neither class: it is a constant of the action,
yet the quantity that actually controls the geometry, $\hat\alpha=\alpha/M^{2}$,
increases at every iteration because the mass decreases. The correction is thus
self-amplifying, and the effective potentials, the horizon, and the ergosphere must be
recomputed at each step with a progressively larger $\hat\alpha$.

This running has direct numerical consequences. A larger initial coupling lowers the
extremal spin and contracts the ergosphere, so the extractable reservoir shrinks and the
sequence stops after fewer decays. At $\hat r=1.2$ the number of admissible events drops
from eight in the Kerr limit to a single decay at $\hat\alpha_{0}=0.0006$, and the
process is forbidden outright for $\hat\alpha_{0}\gtrsim0.0012$; moving the decay radius
outward restores it, and the strongly coupled regime that is closed at $\hat r=1.2$
reopens at $\hat r=1.5$. Over the entire parameter range the stopping condition is set by
the incident particle, with the ordering
$\hat a_{\mathrm{min},1}<\hat a_{\mathrm{min},2}<\hat a_{\mathrm{min},0}$ preserved for
every coupling. Unlike the electromagnetic case~\cite{Aliporcharged2026}, where the sign
of $\hat Q\hat q_{0}$ can shift to the escaping fragment, the Gauss--Bonnet term
only rescales the bound and never changes which particle governs the termination.

The two efficiency measures respond differently, and this is our main result. The energy
return on investment falls monotonically with $\hat\alpha_{0}$, while the growth of the
irreducible mass is suppressed relative to Kerr, so that a smaller fraction of the lost
rotational energy is locked into the horizon. The utilization efficiency behaves
non-monotonically. Below a critical value $\hat\alpha_{0,\mathrm{change}}\approx0.001$
the $(\hat\alpha_{0},\hat r)$ plane divides into four regions: a forbidden region near
the horizon, a suppressed region, a bounded window in which the Einstein--Gauss--Bonnet
black hole outperforms Kerr, and a second suppressed region at larger radii. Above
$\hat\alpha_{0,\mathrm{change}}$ the two crossover radii merge and only three regions
survive, with the enhanced window now adjacent to the minimum radius. This
reorganization across $\hat\alpha_{0,\mathrm{change}}$ is a genuine change in the topology
of the efficiency landscape, driven by the coupling alone. It has no counterpart in the
geometries examined so far: in Kerr~\cite{Ruffini2025PRL,Ruffini2025PRR} the rotational
energy is largely converted into irreducible mass, in
Reissner--Nordstr\"om~\cite{HuCaiWang2026} the charge cannot be reduced to zero, and in
the accelerating case~\cite{ZengWang2026} the utilization efficiency can exceed one half,
but in none of them does the efficiency structure itself change character as a parameter
is varied.

Extending the analysis to charged incident and captured particles remains a
natural direction for future work, combining the charge-sign selectivity of
the Kerr--Newman process with the geometric running of $\hat\alpha$ found here.

 \newpage 

\appendix

\section{Repetitive Penrose Process at $\hat{r}=1.5$}
\label{app:r15}

For completeness, we present the iterative extraction sequence for an 
alternative decay radius, $\hat{r}=1.5$, in Table~\ref{tab:combined_alpha_r1.5}. 
Shifting the decay zone further from the horizon generally extends the 
maximum number of permissible decays. Most notably, the strong-coupling 
regime ($\hat{\alpha}_0 = 0.0012$), which was entirely kinematically 
suppressed at $\hat{r}=1.2$ (Table~\ref{tab:combined_alpha_r1.2}), 
successfully initiates at this larger radius.

\begin{table}[htbp]
    \centering
    \setlength{\arrayrulewidth}{0.5pt}
    \renewcommand{\arraystretch}{1.2}
    \scriptsize
    \setlength{\tabcolsep}{3pt}
\resizebox{\textwidth}{!}{%
\begin{tabular}{cccccccccccccc}
\toprule
 $\hat{\alpha}_0$ & $n$ & $M/M_0$ & $\hat{a}_n$ & $\hat{\alpha}_n$ & $E_{\text{extractable}}/M_0$ & $E_{\text{extracted}}/M_0$ & $M_\text{irr}/M_0$ & $\tilde{\mu}_{1,n}$ & $\hat{E}_{1,n}$ & $\hat{p}_{\phi0,n}$ & $\hat{a}_{\min,0,n}$ & $\xi_n$ & $\Xi_n$  \\
\midrule

\multirow{21}{*}{0} & 0 & 1 & 1. & 0. & 0.292893 & 0 & 0.707107 & 0.0218184 & -3.52063 & 2.26795 & 0.94949 & 0 & 0 \\
 & 1 & 0.999232 & 0.997291 & 0. & 0.267144 & 0.000768146 & 0.732087 & 0.021711 & -3.5383 & 2.27596 & 0.94949 & 0.0768146 & 0.0298324 \\
 & 2 & 0.998464 & 0.994597 & 0. & 0.256703 & 0.00153635 & 0.741761 & 0.0216035 & -3.55614 & 2.28408 & 0.94949 & 0.0768173 & 0.0424519 \\
 & 3 & 0.997695 & 0.991918 & 0. & 0.2488 & 0.0023046 & 0.748896 & 0.0214961 & -3.57414 & 2.29232 & 0.94949 & 0.0768199 & 0.0522663 \\
 & 4 & 0.996927 & 0.989254 & 0. & 0.242218 & 0.0030729 & 0.754709 & 0.0213888 & -3.59233 & 2.30067 & 0.94949 & 0.0768225 & 0.060639 \\
 & 5 & 0.996159 & 0.986605 & 0. & 0.236485 & 0.00384126 & 0.759674 & 0.0212815 & -3.61069 & 2.30914 & 0.94949 & 0.0768251 & 0.0680971 \\
 & 6 & 0.99539 & 0.983971 & 0. & 0.231357 & 0.00460966 & 0.764033 & 0.0211741 & -3.62923 & 2.31774 & 0.94949 & 0.0768277 & 0.0749103 \\
 & 7 & 0.994622 & 0.981351 & 0. & 0.226692 & 0.00537812 & 0.76793 & 0.0210668 & -3.64796 & 2.32647 & 0.94949 & 0.0768303 & 0.0812384 \\
 & 8 & 0.993853 & 0.978747 & 0. & 0.222393 & 0.00614663 & 0.771461 & 0.0209595 & -3.66688 & 2.33532 & 0.94949 & 0.0768329 & 0.0871854 \\
 & 9 & 0.993085 & 0.976158 & 0. & 0.218395 & 0.00691519 & 0.77469 & 0.0208522 & -3.686 & 2.34431 & 0.94949 & 0.0768354 & 0.0928234 \\
 & 10 & 0.992316 & 0.973583 & 0. & 0.214651 & 0.0076838 & 0.777666 & 0.0207448 & -3.70532 & 2.35343 & 0.94949 & 0.076838 & 0.0982047 \\
 & 11 & 0.991548 & 0.971024 & 0. & 0.211123 & 0.00845246 & 0.780424 & 0.0206374 & -3.72484 & 2.3627 & 0.94949 & 0.0768406 & 0.103369 \\
 & 12 & 0.990779 & 0.968481 & 0. & 0.207785 & 0.00922117 & 0.782994 & 0.02053 & -3.74458 & 2.3721 & 0.94949 & 0.0768431 & 0.108346 \\
 & 13 & 0.99001 & 0.965952 & 0. & 0.204613 & 0.00998993 & 0.785397 & 0.0204225 & -3.76453 & 2.38166 & 0.94949 & 0.0768456 & 0.113161 \\
 & 14 & 0.989241 & 0.963439 & 0. & 0.201589 & 0.0107587 & 0.787652 & 0.0203149 & -3.78471 & 2.39137 & 0.94949 & 0.0768482 & 0.117834 \\
 & 15 & 0.988472 & 0.960941 & 0. & 0.198698 & 0.0115276 & 0.789774 & 0.0202072 & -3.80512 & 2.40123 & 0.94949 & 0.0768507 & 0.12238 \\
 & 16 & 0.987703 & 0.958459 & 0. & 0.195928 & 0.0122965 & 0.791776 & 0.0200994 & -3.82576 & 2.41126 & 0.94949 & 0.0768532 & 0.126813 \\
 & 17 & 0.986935 & 0.955992 & 0. & 0.193267 & 0.0130655 & 0.793667 & 0.0199916 & -3.84665 & 2.42145 & 0.94949 & 0.0768557 & 0.131145 \\
 & 18 & 0.986166 & 0.953541 & 0. & 0.190707 & 0.0138345 & 0.795459 & 0.0198836 & -3.86779 & 2.43182 & 0.94949 & 0.0768582 & 0.135385 \\
 & 19 & 0.985396 & 0.951105 & 0. & 0.18824 & 0.0146035 & 0.797157 & 0.0197754 & -3.88919 & 2.44236 & 0.94949 & 0.0768607 & 0.139542 \\
 & \textcolor{red}{20} & \textcolor{red}{0.984627} & \textcolor{red}{0.948685} & \textcolor{red}{0.} & \textcolor{red}{0.185858} & \textcolor{red}{0.0153726} & \textcolor{red}{0.798769} & \textcolor{red}{0.0196671} & \textcolor{red}{-3.91086} & \textcolor{red}{2.45308} & \textcolor{red}{0.94949} & \textcolor{red}{0.0768632} & \textcolor{red}{0.143622} \\
\midrule

\multirow{19}{*}{0.0003} & 0 & 1 & 0.970969 & 0.0003 & 0.283057 & 0 & 0.716943 & 0.0208991 & -3.554 & 2.31593 & 0.931047 & 0 & 0 \\
 & 1 & 0.999257 & 0.968345 & 0.0003 & 0.259963 & 0.000742752 & 0.739295 & 0.0207928 & -3.5724 & 2.32467 & 0.931047 & 0.0742752 & 0.0321619 \\
 & 2 & 0.998514 & 0.965737 & 0.000300446 & 0.250229 & 0.00148555 & 0.748285 & 0.0206869 & -3.59074 & 2.33348 & 0.93102 & 0.0742777 & 0.0452534 \\
 & 3 & 0.997772 & 0.963143 & 0.000301341 & 0.24293 & 0.00222837 & 0.754842 & 0.0205814 & -3.60901 & 2.34234 & 0.930964 & 0.0742789 & 0.0555331 \\
 & 4 & 0.997029 & 0.960564 & 0.000302688 & 0.236909 & 0.00297115 & 0.76012 & 0.0204763 & -3.6272 & 2.35125 & 0.930881 & 0.0742788 & 0.0643839 \\
 & 5 & 0.996286 & 0.957999 & 0.000304495 & 0.231717 & 0.00371387 & 0.764569 & 0.0203717 & -3.6453 & 2.36022 & 0.930769 & 0.0742774 & 0.0723395 \\
 & 6 & 0.995544 & 0.955449 & 0.000306769 & 0.227122 & 0.00445648 & 0.768421 & 0.0202674 & -3.66331 & 2.36925 & 0.930628 & 0.0742746 & 0.0796732 \\
 & 7 & 0.994801 & 0.952914 & 0.000309522 & 0.222986 & 0.00519894 & 0.771815 & 0.0201637 & -3.68122 & 2.37832 & 0.930458 & 0.0742705 & 0.0865472 \\
 & 8 & 0.994059 & 0.950393 & 0.000312766 & 0.219219 & 0.00594121 & 0.77484 & 0.0200604 & -3.69901 & 2.38744 & 0.930257 & 0.0742651 & 0.0930674 \\
 & 9 & 0.993317 & 0.947885 & 0.000316515 & 0.215758 & 0.00668324 & 0.777558 & 0.0199576 & -3.71667 & 2.39661 & 0.930025 & 0.0742582 & 0.0993078 \\
 & 10 & 0.992575 & 0.945392 & 0.000320789 & 0.212559 & 0.007425 & 0.780016 & 0.0198553 & -3.73419 & 2.40581 & 0.92976 & 0.07425 & 0.105323 \\
 & 11 & 0.991834 & 0.942913 & 0.000325606 & 0.209587 & 0.00816643 & 0.782246 & 0.0197535 & -3.75155 & 2.41505 & 0.929462 & 0.0742403 & 0.111154 \\
 & 12 & 0.991093 & 0.940447 & 0.00033099 & 0.206817 & 0.0089075 & 0.784276 & 0.0196524 & -3.76872 & 2.42431 & 0.929128 & 0.0742292 & 0.116835 \\
 & 13 & 0.990352 & 0.937995 & 0.000336966 & 0.204227 & 0.00964814 & 0.786125 & 0.0195518 & -3.78569 & 2.43359 & 0.928758 & 0.0742165 & 0.122392 \\
 & 14 & 0.989612 & 0.935556 & 0.000343564 & 0.201802 & 0.0103883 & 0.78781 & 0.0194519 & -3.80243 & 2.44288 & 0.928349 & 0.0742022 & 0.127849 \\
 & 15 & 0.988872 & 0.93313 & 0.000350815 & 0.199529 & 0.011128 & 0.789343 & 0.0193527 & -3.8189 & 2.45218 & 0.927899 & 0.0741864 & 0.133224 \\
 & 16 & 0.988133 & 0.930718 & 0.000358755 & 0.197397 & 0.011867 & 0.790736 & 0.0192543 & -3.83508 & 2.46146 & 0.927407 & 0.0741689 & 0.138537 \\
 & 17 & 0.987395 & 0.928318 & 0.000367424 & 0.195398 & 0.0126054 & 0.791996 & 0.0191567 & -3.85093 & 2.47072 & 0.926869 & 0.0741496 & 0.143802 \\
 & \textcolor{red}{18} & \textcolor{red}{0.986657} & \textcolor{red}{0.92593} & \textcolor{red}{0.000376865} & \textcolor{red}{0.193526} & \textcolor{red}{0.0133432} & \textcolor{red}{0.793131} & \textcolor{red}{0.01906} & \textcolor{red}{-3.86641} & \textcolor{red}{2.47994} & \textcolor{red}{0.926283} & \textcolor{red}{0.0741286} & \textcolor{red}{0.149034} \\
\midrule

\multirow{17}{*}{0.0006} & 0 & 1 & 0.943275 & 0.0006 & 0.269958 & 0 & 0.730042 & 0.0200348 & -3.57975 & 2.36506 & 0.912355 & 0 & 0 \\
 & 1 & 0.999283 & 0.94073 & 0.0006 & 0.2536 & 0.000717195 & 0.745683 & 0.0199295 & -3.59889 & 2.37457 & 0.912355 & 0.0717195 & 0.0438449 \\
 & 2 & 0.998566 & 0.938201 & 0.000600862 & 0.244774 & 0.00143444 & 0.753791 & 0.019825 & -3.61772 & 2.38408 & 0.912301 & 0.0717219 & 0.0569596 \\
 & 3 & 0.997848 & 0.935686 & 0.000602589 & 0.238112 & 0.00215165 & 0.759736 & 0.0197214 & -3.6362 & 2.39358 & 0.912193 & 0.0717217 & 0.0675656 \\
 & 4 & 0.997131 & 0.933186 & 0.000605191 & 0.232639 & 0.00286876 & 0.764492 & 0.0196186 & -3.65434 & 2.40306 & 0.91203 & 0.071719 & 0.0768729 \\
 & 5 & 0.996414 & 0.9307 & 0.000608678 & 0.227957 & 0.00358569 & 0.768458 & 0.0195166 & -3.6721 & 2.41251 & 0.911811 & 0.0717138 & 0.0853717 \\
 & 6 & 0.995698 & 0.928228 & 0.000613067 & 0.223854 & 0.00430236 & 0.771843 & 0.0194156 & -3.68947 & 2.42194 & 0.911535 & 0.071706 & 0.0933198 \\
 & 7 & 0.994981 & 0.92577 & 0.000618376 & 0.220205 & 0.0050187 & 0.774776 & 0.0193156 & -3.70642 & 2.43132 & 0.911202 & 0.0716957 & 0.100874 \\
 & 8 & 0.994265 & 0.923326 & 0.00062463 & 0.216927 & 0.00573461 & 0.777338 & 0.0192165 & -3.72293 & 2.44064 & 0.910809 & 0.0716827 & 0.108138 \\
 & 9 & 0.99355 & 0.920895 & 0.000631856 & 0.213962 & 0.00645003 & 0.779588 & 0.0191186 & -3.73895 & 2.4499 & 0.910355 & 0.071667 & 0.115189 \\
 & 10 & 0.992835 & 0.918476 & 0.000640087 & 0.211269 & 0.00716486 & 0.781566 & 0.0190217 & -3.75445 & 2.45908 & 0.909838 & 0.0716486 & 0.122083 \\
 & 11 & 0.992121 & 0.91607 & 0.000649358 & 0.208817 & 0.00787902 & 0.783304 & 0.018926 & -3.76939 & 2.46817 & 0.909255 & 0.0716275 & 0.128867 \\
 & 12 & 0.991408 & 0.913676 & 0.000659713 & 0.206583 & 0.00859242 & 0.784825 & 0.0188316 & -3.78373 & 2.47714 & 0.908603 & 0.0716035 & 0.135581 \\
 & 13 & 0.990695 & 0.911294 & 0.000671198 & 0.204549 & 0.00930496 & 0.786146 & 0.0187386 & -3.7974 & 2.48598 & 0.907881 & 0.0715766 & 0.142258 \\
 & 14 & 0.989983 & 0.908924 & 0.000683866 & 0.202702 & 0.0100165 & 0.787282 & 0.018647 & -3.81036 & 2.49467 & 0.907083 & 0.0715467 & 0.148932 \\
 & 15 & 0.989273 & 0.906564 & 0.000697774 & 0.201032 & 0.0107271 & 0.788241 & 0.0185569 & -3.82254 & 2.50318 & 0.906206 & 0.0715137 & 0.155633 \\
 & \textcolor{red}{16} & \textcolor{red}{0.988564} & \textcolor{red}{0.904215} & \textcolor{red}{0.000712989} & \textcolor{red}{0.199532} & \textcolor{red}{0.0114364} & \textcolor{red}{0.789031} & \textcolor{red}{0.0184686} & \textcolor{red}{-3.83388} & \textcolor{red}{2.51149} & \textcolor{red}{0.905247} & \textcolor{red}{0.0714775} & \textcolor{red}{0.162391} \\
\midrule

\multirow{3}{*}{0.0012} & 0 & 1 & 0.877287 & 0.0012 & 0.218507 & 0 & 0.781493 & 0.0178429 & -3.73174 & 2.5345 & 0.874144 & 0 & 0 \\
 & 1 & 0.999334 & 0.874984 & 0.0012 & 0.214578 & 0.00066585 & 0.784756 & 0.0177388 & -3.75388 & 2.5468 & 0.874144 & 0.066585 & 0.169486 \\
 & \textcolor{red}{2} & \textcolor{red}{0.998668} & \textcolor{red}{0.872697} & \textcolor{red}{0.0012016} & \textcolor{red}{0.21114} & \textcolor{red}{0.00133174} & \textcolor{red}{0.787528} & \textcolor{red}{0.0176365} & \textcolor{red}{-3.77508} & \textcolor{red}{2.55891} & \textcolor{red}{0.87404} & \textcolor{red}{0.0665871} & \textcolor{red}{0.180784} \\

\bottomrule
\end{tabular}
}
\caption{Repetitive Penrose process for neutral particles at $\hat{r}=1.5$ for various initial $\hat{\alpha}_0$ in rotating black holes in 4D Einstein-Gauss-Bonnet gravity. 
The rows shown in red correspond to the first inadmissible iteration ($n=n_f+1$), for which $\hat a_n<\hat a_{\min,0,n}$. 
They are included only to verify the termination criterion; the physical repetitive Penrose process ends at the preceding row. Fixed parameters as in Table~\ref{tab:combined_alpha_r1.2}.}
\label{tab:combined_alpha_r1.5}
\end{table}


\newpage

\begin{thebibliography}{20}

\bibitem{Penrose1969}
R.~Penrose,
``Gravitational collapse: The role of general relativity,''
Riv.\ Nuovo Cimento \textbf{1}, 252 (1969).

\bibitem{PenroseFloyd1971}
R.~Penrose and R.~M.~Floyd,
``Extraction of rotational energy from a black hole,''
Nature Phys.\ Sci.\ \textbf{229}, 177 (1971).

\bibitem{Bardeen1972}
J.~M.~Bardeen, W.~H.~Press, and S.~A.~Teukolsky,
``Rotating black holes: Locally nonrotating frames, energy extraction,
and scalar synchrotron radiation,''
Astrophys.\ J.\ \textbf{178}, 347 (1972).

\bibitem{Wald1974}
R.~M.~Wald,
``Energy limits on the Penrose process,''
Astrophys.\ J.\ \textbf{191}, 231 (1974).



\bibitem{Denardo1973}
G.~Denardo and R.~Ruffini,
``On the energetics of Reissner--Nordstr\"{o}m geometries,''
Phys.\ Lett.\ B \textbf{45}, 259 (1973).

\bibitem{Ruffini1975}
R.~Ruffini,
``On the energetics of black holes,''
in \textit{Black Holes -- Les Houches 1972},
ed.\ C.~DeWitt and B.~S.~DeWitt
(Gordon and Breach, New York, 1973).


\bibitem{Ruffini2025PRL}
R.~Ruffini, M.~Prakapenia, H.~Quevedo, and S.~Zhang,
``Single versus the repetitive Penrose process in a Kerr black hole,''
Phys.\ Rev.\ Lett.\ \textbf{134}, 081403 (2025),
arXiv:2405.08229 [gr-qc].

\bibitem{Ruffini2025PRR}
R.~Ruffini, C.~L.~Bianco, M.~Prakapenia, H.~Quevedo, J.~A.~Rueda,
and S.~R.~Zhang,
``Role of the irreducible mass in repetitive Penrose energy extraction
processes in a Kerr black hole,''
Phys.\ Rev.\ Res.\ \textbf{7}, 013203 (2025),
arXiv:2405.10459 [gr-qc].






\bibitem{Christodoulou1970}
D.~Christodoulou,
``Reversible and irreversible transformations in black-hole physics,''
Phys.\ Rev.\ Lett.\ \textbf{25}, 1596 (1970).

\bibitem{ChristodoulouRuffini1971}
D.~Christodoulou and R.~Ruffini,
``Reversible transformations of a charged black hole,''
Phys.\ Rev.\ D \textbf{4}, 3552 (1971).

\bibitem{Hawking1971}
S.~W.~Hawking,
``Gravitational radiation from colliding black holes,''
Phys.\ Rev.\ Lett.\ \textbf{26}, 1344 (1971).

\bibitem{Bekenstein1973}
J.~D.~Bekenstein,
``Black holes and entropy,''
Phys.\ Rev.\ D \textbf{7}, 2333 (1973).



\bibitem{HuCaiWang2026}
L.~Hu, R.-G.~Cai, and S.-J.~Wang,
``Third law of repetitive electric Penrose processes,''
Phys. Rev. D 113, L061501 (2026). 


\bibitem{WangZeng2025}
K.~Wang and X.-X.~Zeng,
``Repetitive Penrose process in Kerr--de~Sitter black holes,''
arXiv:2512.05491 [gr-qc] (2025).

\bibitem{ZengWang2026}
X.-X.~Zeng and K.~Wang,
``Repetitive Penrose process in accelerating Kerr black holes,''
arXiv:2601.01414 [gr-qc] (2026).

\bibitem{Aliporcharged2026}
M.~R.~Alipour, S.~N.~Gashti, and M.~A.~S.~Afshar,
``Repetitive Penrose process for charged particles in Kerr-Newman black holes,''
arXiv:2606.30969 (2026).

\bibitem{Zeng2026KZ}
X.~X.~Zeng, D.~P.~Su, and K.~Wang,
``Repetitive Penrose process in Konoplya-Zhidenko rotating non-Kerr black holes,''
Science China Physics, Mechanics and Astronomy 69, 270414 (2026).

\bibitem{Kumar2020EGB}
R.~Kumar and S.~G.~Ghosh,
``Rotating black holes in 4D Einstein--Gauss--Bonnet gravity and its shadow,''
J. Cosmol. Astropart. Phys. 2020(07), 053 (2020).

\end{thebibliography}
\end{document}